\documentclass[11pt]{article}
\usepackage{graphicx}
\usepackage[margin=1.25in]{geometry}
\usepackage[usenames,dvipsnames]{color}
\usepackage{url}
\usepackage[colorlinks = true,
            linkcolor = blue,
            urlcolor  = blue,
            citecolor = blue,
            anchorcolor = blue]{hyperref}

\newcommand\pubnumber{Transcendental Preprint }
\newcommand\pubdate{\today}

\def\Title#1{\begin{center} {\LARGE #1 } \end{center}}
\def\Author#1{\begin{center}{ \sc #1} \end{center}}

\newcommand\pubblock{\rightline{\begin{tabular}{l} \pubnumber\\
         \pubdate \end{tabular}}}
\newenvironment{Abstract}{\begin{quotation} \begin{center}
                       ABSTRACT
     \end{center}\bigskip  }{\end{quotation}}

\def\beq{\begin{equation}}
\def\eeq#1{\label{#1}\end{equation}}
\def\eeqn{\end{equation}}

\newenvironment{Eqnarray}%
   {\arraycolsep 0.14em\begin{eqnarray}}{\end{eqnarray}}
\def\beqa{\begin{Eqnarray}}
\def\eeqa#1{\label{#1}\end{Eqnarray}}
\def\eeqan{\end{Eqnarray}}

\let\bar=\overbar

\def\lsim{\mathrel{\raise.3ex\hbox{$<$\kern-.75em\lower1ex\hbox{$\sim$}}}}
\def\gsim{\mathrel{\raise.3ex\hbox{$>$\kern-.75em\lower1ex\hbox{$\sim$}}}}

\def\del{\partial}
\def\Dslash{\not{\hbox{\kern-4pt $D$}}}
\def\dslash{\not{\hbox{\kern-2pt $\del$}}}
\def\pslash{\not{\hbox{\kern-2pt $p$}}}
\def\ETmiss{\not{\hbox{\kern-4pt $E$}}_T}

\def\Dlr{\mathrel{\raise1.5ex\hbox{$\leftrightarrow$\kern-1em\lower1.5ex\hbox{$D$}}}}

\def\MSB{{\bar{M \kern -2pt S}}}
\def\msb{{\bar{\scriptsize M \kern -1pt S}}}

\def\drb{{\bar{\scriptsize D \kern -1pt R}}}

\newcommand\snowmass{\begin{center}\rule[-0.2in]{\hsize}{0.01in}\\\rule{\hsize}{0.01in}\\
\vskip 0.1in Submitted to the  Proceedings of the US Community Study\\ 
on the Future of Particle Physics (Snowmass 2021)\\ 
\rule{\hsize}{0.01in}\\\rule[+0.2in]{\hsize}{0.01in} \end{center}}

\begin{document}

\pubblock

\Title{Big Industry Engagement to Benefit HEP: Microelectronics Support from Large CAD Companies}

\bigskip 

\Author{\textbf{Gabriella Carini$^1$, Marcel Demarteau$^2$, Peter Denes$^3$, Angelo Dragone$^4$, Farah Fahim$^5$, Carl Grace$^3$, Shaorui Li$^5\ast$, F Mitch Newcomer$^6$, Brianna Yi$^7$\\}
$^1$Brookhaven National Laboratory \\
$^2$Oak Ridge National Laboratory \\
$^3$Lawrence Berkeley National Laboratory \\
$^4$SLAC National Accelerator Laboratory \\
$^5$Fermi National Accelerator Laboratory \\
$^6$University of Pennsylvania \\
$^7$Pacific Northwest National Laboratory \\
$^\ast$E-mail: shaorui@fnal.gov  }

\medskip


\medskip

\begin{Abstract}
Microelectronics development is critical to a wide number of DOE projects and mission space. Creating Helpful Incentives to Produce Semiconductors (CHIPS) and manufacturing Application Specific Integrated Circuits (ASIC) are important to DOE, so the infrastructure that allows DOE to carry out its mission needs to exist. This paper discusses the current initiatives and recommends a business model to build an ecosystem for microelectronics design for DOE which includes three main building blocks: the Computer Aided Design (CAD) – Electronic Design Automation (EDA) design tools, basic design IPs, and access to semiconductor fabrication facilities.
\end{Abstract}

\snowmass
\def\thefootnote{\fnsymbol{footnote}}
\setcounter{footnote}{0}

\pagebreak
\tableofcontents

\pagebreak

\section*{Executive Summary}
Microelectronics development is critical to a wide number of DOE projects and mission space. Creating Helpful Incentives to Produce Semiconductors (CHIPS) and manufacturing Application Specific Integrated Circuits (ASIC) are important to DOE so the infrastructure that allows DOE to carry out its mission needs to exist. An ecosystem for microelectronics design for DOE includes three main building blocks: the Computer Aided Design (CAD) – Electronic Design Automation (EDA) design tools, basic design IPs (i.e., building blocks for ASIC designs), and access to semiconductor fabrication facilities (i.e., foundries).

The cost of ASIC design licenses has steadily risen over years. The type of licenses required for new projects in advanced technologies has also contributed to the increase of costs. The increased number of projects has resulted in a $\sim$3x growth of the microelectronics teams (e.g., Fermilab), but not a 3x increase in license costs. We work hard to be efficient in sharing licenses among multiple engineers. Collective bargaining will provide economies of scale. The European community has established Europractice IC Service by engaging the large CAD companies to provide research licenses. We are investigating a similar possibility to promote US competitiveness. 

\section{Motivation}
The development of modern microelectronics is a highly sophisticated and complex endeavor. There are few companies that have the capabilities to take on this challenge that requires a range of deep expertise in device and circuit performance and their limitations, as well as sophisticated CAD-EDA tools. Design of ASICs for DOE extreme environments, such as high ionization radiation or cryogenic temperatures, does not have a significant commercial market to engage large companies in developing the required solutions. Currently DOE national labs with academic and other collaborators spearhead the development of chip design for next generation instrumentation required by the DOE mission.  

The major bottleneck is the access to low-cost, high-volume microelectronics CAD tools. The increasing complexity of designing in smaller geometry nodes has led to complicated and expensive licensing frameworks, often with one license being shared among multiple (e.g., more than 10) designers. As the technology node scales, even small designs take significantly long completion time and are harder to debug with limited licenses. Moreover, the existing framework is not suitable for collaborative development especially in joint teams composed of groups from national labs, international labs, universities, and small businesses. Finally, legal clauses for standard IP access are independently renegotiated by each DOE lab, resulting in significant delays and different outcomes.

\section{Goal}
\label{sec:goal}
Overcoming the obstacles to cost-effective tool access and establishing an umbrella framework for collaborative development across multiple institutions will allow the growth of existing groups and help create new groups to support federally funded scientific research. CAD-EDA licenses are intertwined with CAD design tool flows and base design IPs which are tied to a specific technology from a specific foundry. The overall long-term goal is to create a “DOE microelectronics: Tools and IP access program” by establishing a mechanism for integrated and collaborative access to tools, IPs, and semiconductor foundries. 

We aim to create an inclusive and expandable program by engaging not only with the three major CAD-EDA vendors, namely Cadence, Synopsys and Siemens but others such as Ansys, Silvaco, Keysight, etc. Additionally, design IP vendors such as ARM, TSMC etc. will also be added. Tool interoperability is always a challenge but essential for high performance design solutions, hence it is important to engage with multiple vendors. The capability of plugging-in open-source design tools and platforms are especially important for extreme environments since they provide community-driven solutions, beyond the needs of the commercial sector. 

Several main requirements and challenges include:  
\begin{itemize}
     \item Access to low-cost, high-volume research licenses: currently CAD companies categorize licenses as educational (often free of charge) or professional licenses. It's important to create a middle ground for research licenses.
     \item Enable collaborative research across DOE Labs, universities, and international partners while preserving the possibility of commercial private partnership.
     \item A collective strategy for negotiation of terms and conditions, instead of each lab individually negotiating which often takes several months of effort.
     \item Labs often have different requirements. Hence a one-size-fits-all model is not sustainable.
\end{itemize}

\section{Status of Current Initiatives}
\subsection{Meetings with CAD Companies}
DOE Office of High Energy Physics hosted initial meetings with major CAD and EDA tool companies including Ansys, Cadence, Google, Keysight, Siemens, and Synopsys in 2021. Business collaboration models and recommendations are presented and discussed. Here's a summary of major recommendations:
\begin{itemize}
     \item Consider the concept of a DOE Collaborative Innovation Hub scoped for cooperative across the team shared access to CAD/EDA tools, training, and support.
     \item Establish a dedicated cloud-based communal participation between academia, DOE national labs, and CAD/EDA companies.
     \item Leverage successful solution frameworks (e.g. DARPA Innovation Package, Europractice IC Service, DOD Cloud Access Rights) through the efficiencies of shared access.
     \item Incorporate some aspects of CAD/EDA companies' academia policies for research projects at national labs.
     \item Leverage the academic network and cultivate talents to advance and promote innovations in semiconductor technologies.
     \item The solutions need to keep intact the premise of CAD/EDA companies' contributions, with special arrangements for commercializing research results.
     \item Build an Ecosystem including the CAD/EDA tools, available technologies, vendor support, and business models. 
\end{itemize}

\subsection{DARPA Conversations} 

DARPA Toolbox is an Agency-wide effort to provide open licensing opportunities with commercial technology vendors to the researchers behind DARPA programs. DARPA Toolbox provides easy, low-cost, scalable access to state-of-the-art tools and intellectual property (IP) under predictable legal terms and streamlined acquisition procedures. The goal is to reduce performer reliance on low-quality, low-cost tools and IP that increase execution risks and complicate post-DARPA transitions \cite{DarpaToolbox}.

The DARPA team has created a mutually beneficial value proposition with the industry, exploiting the DARPA brand association by allowing highly visible public announcements. The vendors also benefit from adapting tools for cutting edge research programs which have high potential of becoming mainstream solutions in the long term.

They have created a two-tier system where DARPA negotiates low-cost uniform pricing with vendors by utilizing a light weight contract. Then DARPA performers can choose any vendors from the toolbox to buy a subset of licenses to create their own package required for their program. The Terms and Conditions are individually negotiated by the participants with guaranteed pre-determined low-cost prices.
DARPA is currently also looking at deploying an ``all of federal government" approach to broaden the impact and serve national interest. 
DOE ICPT team is engaged with DARPA to learn from their experience while tailoring the concept to DOE needs and programs.

\subsection{ICPT Engagement}
The Integrated Contractor Purchasing Team's (ICPT) objective is to provide ``Guiding Principles" that incorporate best practices in the selection of both commodities/services and suppliers. The strategic selection of commodities/services and suppliers will enhance long-term relationships with suppliers and maximize process and cost savings within the DOE Complex. The ICPT philosophy is to 1) leverage the buying power of the DOE Complex to achieve the most favorable purchasing arrangements and pricing; 2) avoid unnecessary duplication in the acquisition process; 3) establish long-term relationships with quality suppliers, to optimize the number of suppliers per commodity as much as practicable;, and 4) primarily focus on the acquisition of commercial off-the-shelf commodities and commercial services through small, small disadvantaged, woman owned, Hubzone, Veteran-Owned, Service Disabled Veteran-Owned or other minority business enterprises.

To facilitate microelectronics support for DOE Complex Wide Site and Facility Contractors, ICPT aims to start the process with three strategic agreements to standardize the CAD tool procurement and contracts across DOE. ICPT hopes to negotiate general terms and conditions and fixed pricing discounts with the CAD vendors. Individual DOE M\&O Contractors will be able to select specific titles based upon their individual site needs from the portfolio offered by the CAD companies under their ICPT Agreement. To date, ICPT has initiated conversations with three major CAD companies. In addition, ICPT has been working actively with procurement professionals across DOE complex. Mutual benefits to DOE complex and to the CAD companies to support microelectronics are expected, through streamlining the cumbersome negotiation and procurement processes in the past.

In the long term, a dynamic process of adding new CAD tool vendors, IP vendors and Foundries will also be established.

\section{Deliverable}
\label{sec:deliv}
A centralized business model and legal framework negotiated between DOE contracting and CAD tool/IP vendors, with input from the national labs needs to be developed. This model then becomes the basis for engaging and pre-negotiating overall costs and terms and conditions with the vendors participating in the program. Each laboratory and its collaborators can then independently procure CAD tools and IP based on their individual requirements. The two-step system helps us provide economies of scale for the DOE microelectronics program, while enabling the labs to select tools most applicable to their team.

\section{Business Model Requirements}
\label{sec:bm}
A slightly different business model from the DARPA Toolbox is required. Unlike DARPA programs, DOE projects often require production volumes for large-scale detectors. Higher volume licenses are required to support national lab teams. Therefore, it would be desirable if legal concerns are sorted out at the DOE level and equally applicable to all DOE labs. The following requirements are recommended for this centralized business model:
\begin{itemize}
\item Include all aspects of the design chain - CAD licenses, design flows and design IP, support, and training. 
\item Conducive to collaboration between national labs and universities pursuing federally funded research projects. Ability to engage in global partnerships for international scientific research. 
\item Model-based on research licenses to develop basic science experiments. But also enable technology transfer, work with commercial entities from startups to large established businesses with an additional fee to the vendor or low cost professional licenses. 
\item An ecosystem that allows all CAD vendors and open-source tool developers to participate, with an overlook to a broader engagement beyond just chip design to enable system engineering. 
\end{itemize}

\section{Mutual Impacts between HEP and Microelectronics Industry}
\label{sec:mutual benefits}
The DOE microelectronics development is miniature compared to the commercial microelectronics industry. But the industrial needs for innovative microelectronics for extreme environments (e.g., cryogenic operation, high ionization radiation) and the associated risk mitigation continues to increase, where DOE HEP microelectronics development has spearheaded for the past years. One major driving force is the pursue of quantum supremacy, including quantum computing, quantum sensing, and secure long-distance quantum communications. Innovations and proof of concepts originated from the national labs are benefiting the microelectronics industry. Specialized cryogenic microelectronics design and integration techniques are adopted by the industry. Facilities at the national labs are also of growing interest to commercial partners. 

To CAD-EDA companies, DOE microelectronics teams have been a reliable resource to provide feedback on novel applications of the advanced CAD-EDA tools needed by DOE HEP projects. The close collaboration of national labs with the academic network also cultivate talents to advance and promote the workforce for the microelectronics industry.      








\bibliographystyle{JHEP}


\end{document}